\documentclass[useAMS,usenatbib]{mn2e}
\usepackage{graphicx}
\usepackage{txfonts}

\title[IGR J16351-5806: another close by Compton-thick AGN]
 {IGR J16351-5806: another close by Compton-thick AGN}
\author[A. Malizia et al.]
{A.~Malizia,$^1$ L.~Bassani,$^1$  F.~Panessa,$^2$ A.~De Rosa $^2$, A~J.~Bird$^3$ \\
$^1$ IASF/INAF, via Gobetti 101, I-40129 Bologna, Italy\\
$^2$ IASF/INAF, via del Fosso del Cavaliere 100, I-00133 Roma, Italy \\
$^3$ School of Physics and Astronomy, University of Southampton, SO17 1BJ,
Southampton, U.K.}

\begin{document}

\date{}

\pagerange{\pageref{firstpage}--\pageref{lastpage}} \pubyear{2007}
\maketitle
\label{firstpage}
\begin{abstract}
  
  IGR J16351-5806 has been associated with the Seyfert 2 galaxy ESO
  137-G34, having been first reported as a high energy emitter in the
  third \emph{INTEGRAL/IBIS} survey. Using a new diagnostic tool based
  on X-ray column density measurements {\it vs} softness ratios
  (F$_{2-10~keV}$/F$_{20-100~keV}$), Malizia et al. (2007) identified
  this source as a candidate Compton thick AGN. In the present work we
  have analysed combined \emph{XMM-Newton} and \emph{INTEGRAL} data of
  IGR J16351-5806 in order to study its broad band spectrum and
  investigate its Compton thick nature. The prominent K$\alpha$
  fluorescence line around 6.4 keV (EW $>$ 1 keV) together with a flat
  2-10 keV spectrum immediately point to a highly obscured source.
  The overall spectrum can be interpreted in terms of a transmission
  scenario where some of the high energy radiation is able to
  penetrate through the thick absorption and be observed together with
  its reflection from the surface of the torus. A good fit is also
  obtained using a pure reflection spectrum; in this case the primary
  continuum is totally depressed and only its reflection is observed.
  An alternative possibility is that of a complex absorption, where
  two layers of absorbing matter each partially covering the central
  nucleus are present in IGR J16351-5806. All three scenarios are
  compatible from a statistical viewpoint and provide reasonable AGN
  spectral parameters; more importantly all point to a source with an
  absorbing column greater than 1.5 $\times$ 10$^{24}$ cm$^{-2}$, i.e.
  to a Compton thick AGN.  Because of this heavy obscuration, some
  extra components which would otherwise be hidden are able to emerge
  at low energies and can be studied; a thermal component with kT in
  the range 0.6-0.7 keV and free metal abundance is statistically
  requied in all three scenarios while a scattered power law is only
  present in the pure reflection model.  By providing strong evidence
  for the Compton thick nature of IGR J16351-5806, we indirectly
  confirm the validity of the Malizia et al. diagnostic diagram.

\end{abstract}


\section{Introduction}

Compton thick AGN are by definition those in which the X-ray obscuring
matter has a column density equal to, or larger than, the inverse of
the Thomson cross-section (N$_{H} $ $\ge$ $\sigma_{T}^{-1}$ =1.5
$\times$ 10$^{24}$ cm$^{-2}$).  The cross-sections for Compton
scattering and photoelectric absorption have approximately the same
value for energies of the order of 10 keV, which incidentally
coincides with the lower energy threshold of hard X-ray telescopes
like \emph{INTEGRAL/IBIS} and \emph{Swift/BAT}.  For columns in the
range 10$^{24}$-10$^{25}$ cm$^{-2}$ the nuclear radiation is still
visible above 10 keV and the source is called mildly Compton thick.
For higher values of N$_{H}$, the entire high energy spectrum is
down-scattered by Compton recoil and hence depressed over the entire
X/gamma-ray band.  Generally indirect arguments allow determination of
the Compton thick nature of an AGN, such as the presence of a strong
iron K line complex at 6.4 - 7 keV and the characteristic reflection
spectrum. The presence of Compton thick matter can also be inferred
through the ratio of isotropic versus anisotropic emission; primarily
the L$_{X}$/L$_{[OIII]}$ ratio\footnote{Where the L$_{[OIII]}$ is
  corrected for reddening in the host galaxy by means of the
  H$_{\alpha}$/H$_{\beta}$ Balmer decrement} has been used in the
literature to recognise Compton thick sources (Bassani et al.  1999
and Panessa and Bassani 2002).  There are various reasons why these
AGN are important observational targets. They are a key ingredient of
the Cosmic X-ray Background (CXB) and, through their
absorbed/re-emitted radiation, also of the IR one. The heavy
obscuration present in these objects allows some spectral components
which are otherwise usually hidden by the nuclear emission to emerge
and be studied in some detail. Despite their importance, studies of
Compton thick AGN have so far been hampered by the lack of high energy
coverage in X-ray missions and subsequent difficulties in performing
surveys above 10 keV. This situation has been changed by the advent of
satellites like \emph{INTEGRAL} and \emph{Swift}, which are surveying
a large fraction of the sky with good sensitivity and arcminute
location. As a result, a number of Compton thick candidates are now
emerging and can be the subject of more in-depth studies (Ueda et al.
2007, Comastri et al. 2007, Sazonov et al. 2008).  In particular,
Malizia et al. (2007) studying a sample of 34 \emph{INTEGRAL} AGN in
the X-ray band, proposed a new diagnostic tool which allows the
isolation of Compton thick AGN candidates: this relies on measurement
of the source X-ray column density and softness ratio defined as
F$_{2-10~keV}$/F$_{20-100~keV}$.  Using this tool, Malizia and
co-workers suggested 3 INTEGRAL-detected AGN as Compton thick
candidates (IGR J14175-4641, IGR J16351-5806 and Swift J0601.9-8636)
based on the apparent absence of an absorbing column density
accompanied by a very low softness ratio. One of these sources, Swift
J0601.9-8636, has subsequently been observed by Suzaku (Ueda et al.
2007) and found to be a Compton thick AGN: the Suzaku spectrum
revealed a heavily absorbed power law component with a column density
of N$_{H}$ $\sim$ 10$^{24}$ cm$^{-2}$ and an intense reflection
component with a solid angle $\ge$ 2$\pi$ from a cold optically thick
medium.  In this paper we present \emph{XMM-Newton} and
\emph{INTEGRAL} data of IGR J16351-5806 which provide strong evidence
for the Compton-thick nature of also this object and indirectly
confirm the validity of the diagnostic diagram proposed by Malizia et
al.

\section{IGR J16351-5806 = ESO 137-G34}

IGR J16351-5806 was first reported as a hard X-ray emitter by Bird et
al. (2007) and subsequently confirmed by Krivonos et al. (2007).  Here
and in the following, we use data collected for the 3rd IBIS survey
(Bird et al. 2007), which in this region correspond to a total
exposure of $\sim $1.4\thinspace Ms. A clear excess is present in the
18-60 keV IBIS map with a significance of $\sim$8.3$\sigma$ at a
position corresponding to RA(2000)=16$^{\rm h}$35$^{\rm m}$11.04$^{\rm
  s}$ and Dec(2000)=-58$^{\circ}$05$^{\prime}$24.0$^{\prime\prime}$
and with an associated uncertainty of 3.1$^{\prime}$ (90\% confidence
level). Our position is compatible with that reported by Krivonos et
al (2007) (note that their 2.1$^{\prime}$ error radius corresponds to
68\% confidence level).  The source position was then reduced to
arcsec accuracy by means of \emph{Swift}/XRT observations which
located the source at RA(J2000)=16$^{\rm h}$35$^{\rm m}$13.17$^{\rm s}$, 
Dec(J2000)=-58$^{\circ}$04$^{\prime}$49.68$^{\prime\prime}$
(5.1 arcsec uncertainty) and confirmed the association with the Seyfert 2 
galaxy ESO 137-34 at z = 0.009 (Malizia et al. 2007, Landi et al. 2007).\\
ESO 137-G34 is an S0/a galaxy which in HST (Hubble Space Telescope) images displays a 
long dust lane east of the nucleus, running northwest-southeast, parallel to 
the photometric major axis of the galaxy and a second dust lane perpendicular 
to the first one (Ferruit et al. 2000); moreover in the HST [O III]/[N II] + H$\alpha$ map, 
the morphology of the central regions appears biconical.\\
The source was  observed in the radio waveband with the ATCA telescope at 3.5cm: 
it has a flux of 9.6 mJy and displays a complex morphology consisting of three 
knots aligned along a position angle of -40 degrees (Morganti et al. 1999). 
The radio emission is coincident with the inner part of the line-emitting gas.\\
The source has been studied for the first time in X-rays by Malizia et
al. (2007) using \emph{Swift/XRT} observations, and its X-ray spectrum
was broadly defined as being a flat ($\Gamma$= 1.6) power law with no
absorption in excess of the Galactic value.  The 2-10 keV observed
flux is 3 $\times$ 10$^{-13}$ erg cm$^{-2}$ s$^{-1}$ almost a factor
of 100 weaker than the 20-100 keV flux measured by \emph{INTEGRAL} (2
$\times$ 10$^{-11}$ erg cm$^{-2}$ s$^{-1}$). IGR J116351-5806 was
serendipitously observed by \emph{XMM-Newton} on February 13${\rm
  th}$, 2006 and these X-ray data together with the \emph{INTEGRAL}
ones are used in this work to study the broad band (0.5-100 keV)
spectrum of the source. This analysis provides evidence for the
presence of features typical of a Compton thick AGN.

\begin{figure}
\centering
\includegraphics[angle=-90,width=0.8\linewidth]{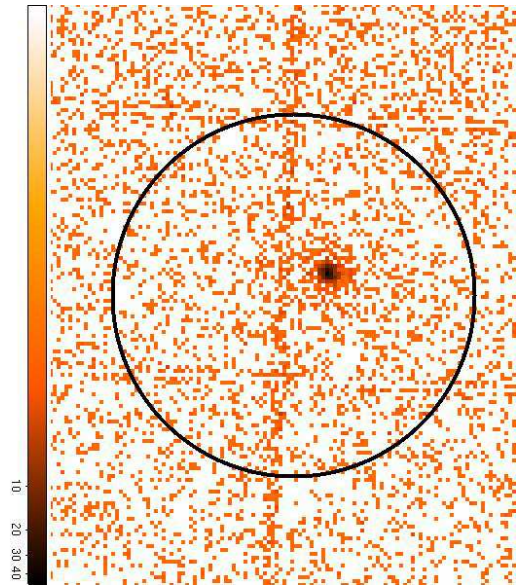}
\caption{\emph {XMM-Newton} pn 4-10 keV image of IGR J16351-5806 with the 3.1 arcmin
INTEGRAL error box
superimposed}
\label{fig2}
\end{figure}

\section{data analysis}

\emph{XMM-Newton} EPIC-pn (Struder et al. 2001) data were reprocessed
using the \emph{XMM} Standard Analysis Software (SAS) version 7.1.2
employing the latest available calibration files. Only patterns
corresponding to single and double events (PATTERN$\leq$4) were taken
into account; the standard selection filter FLAG=0 was applied. The
observation has been filtered for periods of high background and the
resulting exposure is 17 ks.  Source counts were extracted from
circular region of radius 25$^{\prime\prime}$ centred on the source in
order to exclude the extended emission associated to the galaxy;
background spectra were extracted from circular regions close to the
source, or from source-free regions of 80$^{\prime\prime}$ radius. The
ancillary response matrices (ARFs) and the detector response matrices
(RMFs) were generated using the \emph{XMM} SAS tasks \emph{arfgen} and
\emph{rmfgen}; spectral channels were rebinned in order to achieve a
minimum of 20 counts for each bin.  Figure 1 shows the 4-10 keV pn
image of the region surrounding IGR J16351-5806 with the
\emph{INTEGRAL} error box superimposed;
it clearly shows that the AGN is the only source of hard X-ray photons 
and hence the counterpart of the  \emph{INTEGRAL} source. Inspection of 
the XMM light curve provides no evidence for variability over the observing period, 
so all data were used for the spectral analysis. \\
The \emph{INTEGRAL} data reported here consist of several pointings
performed by the \emph{IBIS/ISGRI} (Ubertini et al. 2003, Lebrun et
al. 2003) instrument between revolution 12 and 429, i.e. the period
from launch to the end of April 2006. \emph{ISGRI} images for each
available pointing were generated in various energy bands using the
ISDC offline scientific analysis software OSA (Goldwurm et al. 2003)
version 5.1. Count rates at the position of the source were extracted
from individual images in order to provide light curves in various
energy bands; from these light curves, average fluxes were then
estimated and combined to produce an average source spectrum in the
20-110 keV band (see Bird et al. 2007 for details).

\begin{figure}
\centering
\includegraphics[angle=-90,width=0.8\linewidth]{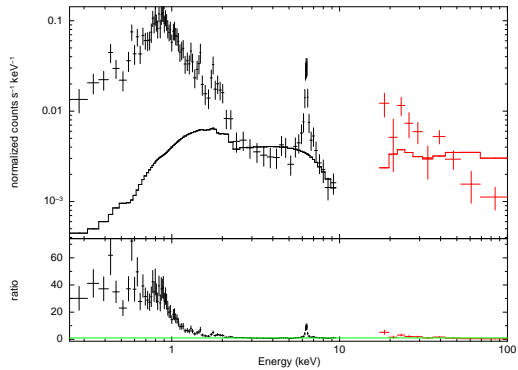}
\caption{ X-ray and Gamma ray data compared to the simple power law used to fit the
2-10 keV data (see text to details); low energy data are from \emph{XMM-Newton} EPIC-pn instrument 
while high energy points are from \emph{INTEGRAL}\emph{IBIS/ISGRI} telescope  }
\label{fig2}
\end{figure}

The \emph{XMM-Newton} data have been fitted together with the
\emph{INTEGRAL} ones using \texttt{XSPEC} v.12.3.1; errors are quoted
at 90\% confidence level for one parameter of interest
($\Delta\chi^2$=2.71). In the fitting procedure, a cross-calibration
constant, \emph{C}, between X-ray and gamma-ray data has been
introduced and fixed to 1 since generally a good agreement between the
two instruments has been found in previous works regarding various
AGNs (De Rosa et al. 2008, Panessa et al. 2008, Molina et al. 2008).
In the following analysis, we always include the Galactic column
density which in the direction of IGR J16351-5806 is
N$_{H_{Gal}}$=2.47 $\times$ 10$^{21}$ cm$^{-2}$
(Dickey \& Lockman  1990).\\
A simple absorbed power law model fails to reproduce the broad band
spectrum of IGR J16351-5806: it gives a steep photon index
($\Gamma$=2.5), no absorption and a $\chi^{2}$ of 607.11 for 93 d.o.f.
However if only the \emph{XMM} data are considered over the 2-10 keV
band, the photon index is much flatter ($\Gamma \ll$ 1), there still
is no evidence of absorption in excess to the Galactic value, and the
observed 2-10 keV flux is $\sim$ 5 $\times$ 10$^{-13}$ erg cm$^{-2}$
s$^{-1}$. This spectrum is similar to that first observed in the
prototype Compton thick AGN NGC 1068 (Matt et al. 2000). Although the
\emph{XMM} spectrum is flatter than the \emph{XRT} one (probably due
to the different energy band used), the X-ray fluxes are compatible,
suggesting no significant variation over one year time scale.  The use
of the \emph{XMM} fit over a broader band (0.2-100 keV, see figure 2)
helps in identifying extra features which contribute to the overall
emission.  Indeed, inspection of figure 2 indicates the presence of a
strong soft excess and a prominent
line at around 6.4 keV. These features, together with the flat photon index, strongly resemble a Compton thick AGN.\\
The shape of the emission below 2 keV and the presence of emission
lines clearly indicate the presence of gas photoionized by the central
engine of the AGN (see discussion). This component is often present in
Compton thick AGN (Guainazzi et al. 1999, Guainazzi et al. 2005)
where, thanks to the heavy absorption, it emerges and can be well
studied. It is generally fitted in \texttt{XSPEC} with a
\texttt{mekal} model having free metal abundance.
Indeed the addition of a \texttt{mekal} model improves the fit ($\Delta\chi^2$=377 for 2 d.o.f.) 
and gives a gas temperature kT of 0.64$^{+0.04}_{-0.05}$ keV and, as already found in other absorbed AGN, 
a low metal abundance value A$_{Z}$ of 0.07$^{+0.03}_{-0.02}$, still leaving an extremely flat power law 
continuum ($\Gamma$$\sim$0.3). Although this model yields an adequate fit to our spectrum, 
it is worth noting that it may be an oversimplified parameterization of the data. High resolution 
spectroscopy of nearby Seyfert 2s have, in fact, demonstrated that the soft X-ray emission is often 
dominated by emission lines with negligible contribution by an underlying continuum. Blending of 
these emission lines in the EPIC spectra can mimic a continuum emission (Iwasawa et al. 2003).  
As our main goal is to prove the Compton thick nature of IGR J16351-5806, the uncertainties 
induced by a purely phenomenological modelling of the soft excess will not substantially affect 
the core results of our paper.\\
Another significant improvement ($\Delta\chi^2$=103 for 3 d.o.f.)  is obtained when we introduce 
the $K\alpha$ iron fluorescence emission line: the energy of the line is found to be at 
6.39 $\pm$ 0.02 keV with an equivalent width of about 2 keV. The intrinsic width of the line is 
narrow (0.09$\pm$0.04) and for simplicity has been frozen to the observed value in subsequent fits. 
The flat slope of the power law ($\sim$0.3) as well as the strength of the line at 6.4 keV 
(EW $\ge$ 1 keV) and the lack of absorption again suggest that the source is 
 highly absorbed and likely to be in the Compton thick regime (Matt et al. 2000).\\
Two models are generally used to account for the soft gamma-ray  continuum of AGN in this regime:

\begin{enumerate}
\item a transmission model more appropriate to mildly Compton thick AGN; in this scenario 
the power law is absorbed by a column density in the range 10$^{24}$ -- $<$ 10 $^{25}$ and a Compton 
reflection component is often present and added to the fit (model \texttt{MEKAL + WA * PO + ZGA + PEXRAV} 
in \texttt{XSPEC})\footnote{this is the model also adopted by Ueda et al. (2007) 
to describe the Suzaku spectrum of Swift J0601.9-8636}
\item  a pure reflection model more typical of Compton thick AGN with N$_{H}$ $>$ 10$^{25}$ cm$^{-2}$; 
in this case the absorbed power law is no longer detected as it is totally absorbed and only the 
reflection component is visible (model \texttt{MEKAL + PEXRAV + ZGA} in \texttt{XSPEC}).
\end{enumerate}

In order to evaluate the contribution of the Compton reflection component, we used the code 
developed by  Magdziarz \& Zdziarski (1995), where we fixed the cosine of the inclination angle 
to 0.5 and the cut-off energy at 1 MeV.\\
In the transmission scenario  we find a column density of N$_{H}$ = 4.8 $\times$10$^{24}$ cm$^{-2}$, 
a photon index of 1.8 and a reflection of  $\sim$1 ($\chi^{2}$=98.6/86); it is worth noting that in this 
scenario the reflection component is highly requested as without it the model is unable to reproduce the INTEGRAL data.\\
The reflection model yields a good fit too (89.45/88): the high energy part of the spectrum is now 
represented by a reflection of  0.6 while the  column density is indirectely estimated to be 
N$_{H}$ $>$  10$^{25}$ cm$^{-2}$, i.e. a value typical of reflection dominated Seyfert 2 galaxies. 
In both models, the temperature and abundance of the thermal  component are consistent while 
the iron line mantains an equivalent width around 1.5 keV. Both scenarios are able to explain 
such high values of EW and are therefore equally viable (Ghisellini, Haardt \& Matt 1994). 
Despite the good fits, neither model (transmission or reflection) accounts for the excess 
counts in the range 2-5 keV which are still visible in the residuals and require the introduction 
of an additional component in the form of a power law. This addition is marginally required by the 
data in the transmission model ($\sim$90\%), while it is more significant in the reflection model ($>$99\%). 
If we assume that this extra component is due to scattered emission from the primary continuum, 
then the scattered fraction is around 6\% for the reflection model and 10\%  for the transmission one; 
both values are similar to those typically observed in Seyfert 2 galaxies. If present, this soft component 
may be associated with the radio knots and emission line cones observed by ATCA and HST, or it could be due 
to soft emission from the galaxy and/or starburst  activity in the source.\\
Note however that the addition of this extra component in the transmission model flattens the 
primary continuum ($\Gamma$ = 1.31$^{+0.16}_{-0.17}$), gives a smaller column density 
(N$_{H}$ $\sim$ 9$\times$ 10$^{23}$ cm$^{-2}$) and has more difficulties in explaining 
a large iron line EW. On the basis of these indications, and in view of  the marginal evidence 
in the data of this extra component, we prefer to neglect it in the transmission scenario but keep 
it in the reflection one; the relative best fit parameters for both models
are given in columns 1 and 2 of table 1. \\ 
We also consider an alternative scenario where IGR J16351-5806  is characterised by a complex 
absorption partially covering the central nucleus (\texttt{ MEKAL + PCFABS * PCFABS * PO + ZGA}  
in \texttt{XSPEC}). This model mimics the reflection component and provides a canonical AGN  
spectrum ($\Gamma$=1.8) for the source (see table 1, column 3). The absorption required by the data 
is in the form of two columns (N$_{H_{1}}$ $\sim$ 4 $\times$ 10$^{24}$ cm$^{-2}$ and 
N$_{H_{2}}$ $\sim$ 6 $\times$ 10$^{23}$ cm$^{-2}$), both covering 80-90\% of the source. 
Also in this case the combinations of such high columns are able to explain the observed 
iron line EW (Ghisellini, Haardt \& Matt 1994).\\
Despite being rarely used in the literature to model Compton thick AGNs, this model is 
of interest in view of recent studies on the torus geometry and nature  which strongly indicate that 
this  structure is clumpy and made of dusty clouds that are individually optically thick (Elitzur 2008). 
It is possible that in IGR J16351-5806, the torus is stratified with a Compton thick region which 
progressively decreases in column density moving away from its plane; in this case two layers with 
different absorption (one Compton thick and one Compton thin) are intercepted by our line of sight 
(Risaliti et al. 2005); if this is the case, variations in the column densities are expected as 
the clouds move through these two regions. Besides giving the best $\chi^{2}$ (88.7/85) of all 
3 models used here, the complex absorption scenario is also attractive for its simplicity as it does not 
require additional spectral features such as the reflection and the extra power law components. 
Although further X-ray observations of the source can confirm or deny the presence of a complex 
absorber by measuring variations in the column densities, we tentatively assume this to be the 
more appropriate scenario for this source. The unfolded spectrum of IGR J16351-5806 fitted 
with this model  is shown in figure 3.

\begin{figure}
\centering
\includegraphics[angle=-90,width=0.8\linewidth]{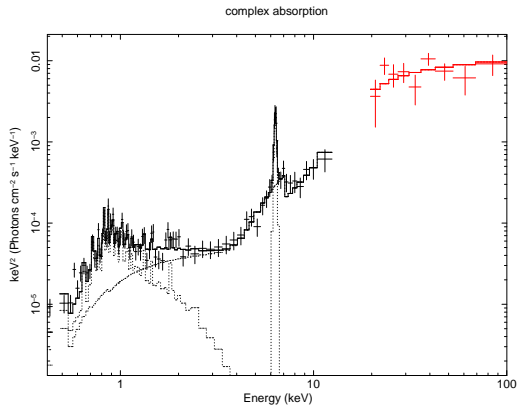}
\caption{X-ray (XMM-Newton) and $\gamma$-ray (INTEGRAL-IBIS) data fitted with the
complex absorption model 
(\texttt{mekal+pcfabs*pcfabs*po+zga})}.
\label{fig2}
\end{figure}

\begin{table}
\begin{center}
\begin{minipage}{140mm}
\centerline{Table 1: Fit parameters}
\label{Table 1: Fit Parameters}
 \renewcommand{\footnoterule}{}
\begin{tabular}{l c  c c}
 \hline\hline
                           & \it{transmission} $^{(a)}$                       &    \it{reflection}$^{(b)}$     &  \it{complex abs}$^{(c)}$             \\
\hline                                                                              
kT                         & 0.69$^{+0.06}_{-0.04}$                   & 0.66$^{+0.06}_{-0.05}$  & 0.62$^{+0.05}_{-0.05}$\\ 
A$_{Z}$                    & 0.05$^{+0.02}_{-0.01}$                                 
                          & 0.31$^{+1.18}_{-0.19}$          & $>$0.06     \\
N$_{H_{1}}$ (cm$^{-2}$)    & 4.8$^{+7.4}_{-1.6}$ $\times$ 10$^{24}$  &    -         
          & 3.7$^{+4.8}_{-1.7} $ $\times$ 10$^{24}$  \\
C$_{f_{1}}$                &         -                                &    -        
           &   0.86$^{+0.09}_{-0.18} $ \\
N$_{H_{2}}$ (cm$^{-2}$)    &          -                               &    -        
           &   5.8$^{+2.2}_{-2.2} $ $\times$ 10$^{23}$  \\
C$_{f_{2}}$                &      -                                   &   -         
           &    0.93$^{+0.08}_{-0.09} $ \\
$\Gamma_{soft}$            &     -                                    &
2.42$^{+0.42}_{-1.60}$  &    -   \\
$\Gamma_{hard}$            &   1.82$^{+0.21}_{-0.27}$                 &
1.33$^{+0.07}_{-0.07}$  &     1.78$^{+0.08}_{-0.07} $ \\
 R                         &  1$^{+0.82}_{-0.82}$                         &
0.59$^{+0.83}_{-0.27}$  &   -   \\
E$_{line}$  (keV)          &  6.39$^{+0.02}_{-0.02}$                  &
6.39$^{+0.02}_{-0.02}$  &  6.39$^{+0.02}_{-0.02}$ \\
EW    (keV)                & 1.5$^{+0.27}_{-0.22}$                    &
1.56$^{+0.28}_{-0.29}$  &  1.36$^{+0.50}_{-0.38}$\\
$\chi^{2}$/dof             &  98.59/86                                & 89.45/86    
           &    88.70/85\\ 
\hline 
F$_{2-10~ keV}$            & \multicolumn{3}{c}{$\sim$ 5 $\times$ 10$^{-13}$
cm$^{-2}$ s$^{-1}$} \\
 F$_{20-100~ keV}$         & \multicolumn{3}{c}{$\sim$ 2 $\times$ 10$^{-11}$
cm$^{-2}$ s$^{-1}$ }\\          
\hline
\hline
\end{tabular}
\end{minipage}
\end{center}
$^{(a)}$  \texttt{MEKAL + WA * PO + ZGA + PEXRAV}\\
$^{(b)}$  \texttt{MEKAL +  PO + PEXRAV + ZGA}\\
$^{(c)}$  \texttt{MEKAL + PCFABS * PCFABS * PO + ZGA}\\
\end{table}

\section{Discussion and Conclusions}
In this work we have analysed the combined \emph{XMM-Newton} and \emph{INTEGRAL-IBIS} data of the Seyfert 2 
galaxy IGR J16351-5806. 
The broad band spectrum of IGR J16351-5806 provides immediate strong indications of its Compton thick nature, 
in particular the presence of a flat 2-10 keV spectrum and  of a prominent emission line around 6.4 keV. 
The overall spectrum can either be interpreted in terms of a transmission scenario where some of the 
high energy radiation is able to penetrate through the thick absorption and be observed together with its 
reflection from the surface of the torus. A good fit is also obtained using a pure reflection spectrum: 
in this case the primary continuum is totally depressed and only its reflection is observed. 
An alternative possibility is that of complex absorption, where two layers of absorbing matter 
each partially covering the central nucleus, are present in IGR J16351-5806. 
All  three scenarios are compatible with the data from a statistical view point, 
provide reasonable AGN spectral parameters, and are capable of producing the observed iron line EW.  
They are therefore indistinguishable using the present data set, although the complex absorption 
model is more attractive for its simplicity and its geometrical implications for the torus that 
are more in line with recent observational and theoretical findings. 
Whether mildly absorbed, totally hidden or partially covered, the source has a column density 
exceeding 1.5 $\times$10$^{24}$ cm$^{-2}$, which  confirms IGR J16351-5806 as a new Compton 
thick AGN detected by INTEGRAL in our neighbourhood.\\
Because of this heavy obscuration, some extra components which would be otherwise hidden, 
are able to emerge at low energies and can be studied;  this includes the presence of thermal 
emission with free metal abundance and possibly of a scattered power law.
The presence in the soft X-ray band of a hot diffuse gas with kT $\sim$0.6-0.7 and A$_{Z}$ $>$0.05 
is in agreement  with observational evidence coming from grating instruments onboard \emph{Chandra} 
and \emph{XMM-Newton} on other Compton thick Seyfert 2 galaxies (Guainazzi et a. 2005): 
within the sample analysed by these authors similar temperature and abundance values were found and 
interpreted in terms of emission from an optically thin, collisionally ionized plasma. 
Although this could also be the interpretation in the case 
of IGR J16351-5806, we note the model used to fit the soft part of the spectrum of IGR J16351-5806 is 
still oversimplified and we need RGS data for a more detailed analysis; unfortunately the source is 
too weak  for good quality RGS data in the present observation and needs to be re-observed with a longer exposure.\\
IGR J16351-5806 has been highlighted previously by Malizia et al. (2007) as one of three possible 
Compton thick objects, using a new diagnostic tool based  on the measurement  of the X-ray column 
density  and softness ratio F$_{2-10~keV}$/F$_{20-100~keV}$.  Of their three candidates, 
SWIFT J0601.9-8636, has been subsequently confirmed as a Compton thick AGN thanks to Suzaku 
observations (Ueda et al. 2007), another, IGR J14175-4641 is very dim and with too limited 
spectral information below 10 keV to be properly studied - we hope to overcome this with a future XMM pointing. 
The third source is IGR J16351-5806 discussed here. 
By proving its Compton thick nature we have also indirectly demonstrated the validity of 
the Malizia et al. (2007) diagnostic diagram and provided an extra tool in the search for such rare AGN.
The use of this diagnostic together with other information will be further tested on a larger 
INTEGRAL AGN sample for which 2-10 keV data are now becoming available.

\section*{Acknowledgements}
We acknowledge ASI  financial and programmatic support via contracts I/008/07/0 and
I/088/06/0. 
The authors would like to thanks the anonymous referee for his/her help in
suggesting the complex absorption
as a possible scenario for IGR J16351-5806

\end{document}